\documentclass{emulateapj}
\usepackage{color}
\newcommand{\lb}{$\cal L_{B}$}
\newcommand{\ls}{$\cal L$}
\newcommand{\ie}{{\it i.e.}}
\newcommand{\eg}{{\it e.g.}}
\newcommand{\alf}{Alfv\'en }
\newcommand{\be}{\begin{equation}} 
\newcommand{\ee}{\end{equation}}
\newcommand{\bea}{\begin{eqnarray}} 
\newcommand{\eea}{\end{eqnarray}}
\newcommand{\mt}[1]{{{#1}}}

\shorttitle{AEI, RWI, diskoseismology and HF-QPO}
\shortauthors{Tagger \& Varni\`ere}

\begin{document}
\title{Accretion-Ejection Instability, MHD Rossby Wave Instability, diskoseismology, and the high-frequency QPO of microquasars}

\author{Michel Tagger }
\affil{Service d'Astrophysique, (UMR AstroParticules et Cosmologie), CEA Saclay \\ 91191 Gif-sur-Yvette, France}
\email{tagger@cea.fr}
\author{Peggy Varni\`ere$^1$}
\affil{LAOG, Universit\'e J. Fourier (UMR5571) France}
\email{peggy.varniere@obs.ujf-grenoble.fr}
\altaffiltext{1}{Department of Physics \& Astronomy, Rochester University, 
Rochester NY 14627-0171 }

\begin{abstract}
We \mt{present a possible explanation for} the high-frequency Quasi-Periodic Oscillations of microquasars by an MHD instability that combines the physics developed, in different contexts, for the Accretion-Ejection Instability, the Rossby-Wave Instability, and the normal modes of diskoseismic models (which rely on the properties of the relativistic rotation curve in the vicinity of the Marginally Stable Orbit). This instability can appear as modes of azimuthal wavenumbers $m=2, 3$, \ldots that have very similar pattern speeds $\omega/m$, while the $m=1$ mode, which would appear as the fundamental of this discrete spectrum, is less unstable. This would readily explain the 2:3 (and sometimes higher) frequency ratio observed between these QPO. \\
These instabilites form eigenmodes, \ie\ standing wave patterns at a constant frequency in the disk; they are strongly unstable, and thus do not need an external excitation mechanism to reach high amplitudes. Furthermore, they have the property that a fraction of the accretion energy can be emitted toward the corona: this would explain that these QPO are seen in a spectral state where Comptonized emission from the corona is always present. \mt{Their existence depends critically on the existence of a magnetic structure, formed by poloidal flux advected in the accretion process, in the central region between the disk and the black hole}.
\end{abstract}
\section{Introduction}
Although many models have attempted to explain the Quasi-Periodic Oscillations (QPO) of X-ray binaries, none of them has gained yet a wide acceptance, and many have been found in contradiction with the observed frequencies and properties of the QPOs. The latest addition to this mystery has emerged in the last few years: it is the observation of high-frequency QPO (HF-QPO) in microquasars, where they are observed at different frequencies in a 2:3 (and sometimes higher) ratio \citep[see \eg][]{McR06}. \\
On the other hand, we have  in recent years \citep{RVT02, VRT02, TVRP04} proposed to explain the low-frequency QPO (LF-QPO), observed when the sources are in different spectral states, by an MHD instability of magnetized accretion disks, the Accretion-Ejection Instability \citep[AEI, ][]{TP99}. Besides producing a frequency in the right range, and other properties which support this explanation, it has the advantage of being a {\em normal mode}, \ie\ to produce long-lived patterns at a quasi-stationary frequency, and of being an {\em instability}, \ie\  to grow spontaneously to high amplitudes if the disk is in a favorable magnetic configuration. The AEI is formed of a combination of spiral and Rossby waves propagating in the disk, and coupled by differential rotation. Its name relates to the fact that, if the disk is covered by a low-density corona, a sizable fraction of the accretion energy extracted from the disk can end up in \alf waves emitted by the instability in the corona, where they might energize a jet \citep{VT02}. This might correspond to the fact that the LF-QPO is seen when the source is in the `low-hard' state, where the disk emission is very faint and dominated by a strong and hard emission from the corona. In this sense, although it exists already in an infinitely thin disk in vacuum, and we are able only to treat a low-density corona as a perturbation,  the AEI must be viewed as a three-dimensional instability of the disk-corona system.\\
In this paper we show that  the HF-QPO, observed in a few microquasars when they are in a Very High (or steep-power-low) state, may be explained by an instability, due to the specific conditions near the Marginally Stable Orbit (MSO) of a black hole disk, and which can be described as a hybrid of the AEI and of a different, but closely related instability: an MHD form of the Rossby-Wave Instability  \citep[RWI:][]{LH78, LLC99}. Or, to present this differently, we show that the modes of diskoseismologic models \citep{NW91} are strongly unstable by the AEI-RWI mechanism. All the modes with azimuthal wavenumber $m\geq 2$ have very close pattern speeds $\omega/m$ and growth rates so that, depending on the conditions, any of them can be expected to dominate, readily producing frequencies in the ratio 2:3:4... On the other hand the $m=1$ mode (which would appear as the fundamental in this discrete spectrum) is always much less unstable. Although the instability exists already in unmagnetized disks, it is much stronger when the disk is threaded by a vertical magnetic field. We have recently shown \citep{TM06} that the RWI could explain the quasi-periodic behaviour observed during the flares of the central black hole of the Galaxy in Sgr A*, when a blob of gas from the environment is captured in the disk at a few to a few tens of Schwarzschild radii. Here on the other hand the instability is due essentially to the properties of the relativistic rotation curve, in the vicinity of the MSO.\\
Our paper is organized as follows: in section \ref{sec:RWI} we will present the basic  physics of Rossby waves, and the conditions that can make them unstable (becoming the RWI) in a differentially rotating disk. We will also present the connection between the RWI and the AEI, which share much of their physics. In section \ref{sec:disko} we will discuss the properties of diskoseismologic modes that exist in the specific conditions near the MSO of relativistic disks, and show that they are unstable by our mechanism. Section \ref{sec:Num} will be devoted to numerical results, first via a linearized analyis and then by non-linear MHD simulations which lead us to discuss the magnetic conditions near the inner radius of the disk. The last section will be devoted to a discussion of these results and their astrophysical implications.
\section{Magnetic Rossby wave and instability}\label{sec:RWI}
\subsection{The \lq classical\rq{} RWI}\label{sec:RWIC}
The RWI has been found and discussed in different contexts of differentially rotating disks. In galactic disks, it was introduced by \cite{LH78}, who called it Negative-Mass Instability. They showed  that a disk presenting an extremum of a certain 
quantity \ls\ (later dubbed vortensity, see below for its definition) was subject to a local instability of Rossby vortices. The 
requirement of an extremum is similar to that giving rise to the Kelvin-Helmholtz (KH) instability of 
sheared flows \citep{DR81}, so we may view this as an application of the KH instability to disks 
in differential rotation; more recently, \cite{LLC99}Ê \citep[see also][]{LFL00,LCW01} developed the theory and numerical simulation for accretion disks and renamed it the Rossby 
Wave Instability (RWI). In the meantime it had been re-discovered in numerical simulations of galactic disks, and the theory rederived,
by \cite{SK91}; \cite{DZ91} also discussed it (as a non-linear instability) in the context of accretion 
disks. \cite{VT06} used it to discuss a possible scenario for accretion and planet formation in the {\lq Dead Zone\rq} of protostellar disks. Finally \cite{TM06} pointed to the MHD form of the instability, and showed that it might explain the quasi-periodic modulation seen during the flares in Sgr~A*.\\
\mt{
In isothermal, unmagnetized disks, the quantity $\cal L$ is the specific vorticity averaged across 
the disk thickness, 
\be
{\cal L}\ = \frac W \Sigma
\label{eq:defL}
\ee
where
\be
W = \left(\vec\nabla \times \vec V\right)_{z}=\frac{\kappa^2}{2\Omega}\;,
\label{eq:defW}
\ee
and $\Sigma$ is the disk's surface density, $\Omega$ its rotation frequency, and $\kappa$ 
the epicyclic frequency, defined by
\begin{equation}
\kappa^2\ =\ 4\Omega^2+2\Omega\Omega'r=\frac{2\Omega}{r}\frac{d}{dr}\left(\Omega r^2\right)\;.
\end{equation}
The prime notes the radial derivative, and one notes that  $\kappa=\Omega$ in keplerian disks. In magnetized disks $\cal L$  is replaced by \citep{TP99}
\begin{equation}
{\cal L_{B}}\ =W\ \frac{\Sigma}{B^{2}}\ .
\end{equation}
}
Although the RWI has been already discussed in the above-mentioned references, we present here its basic physics for the sake of completeness. It results from two essential properties of waves in differentially rotating disks:
\begin{itemize}
\item Rossby waves are most often studied in fluids presenting a solid rotation, but a gradient of vorticity, such as planetary atmospheres: the rotation is that of the planet, while vorticity has a gradient with latitude. They propagate perturbations of vorticity  in the flow. Their  dispersion relation is
\begin{equation}
\omega=\frac{k_{y}W'}{k_{x}^{2}+k_{y}^{2}}
\end{equation}
where $x$ is the direction of the gradient (latitude, or radial in what follows), $y$ its normal (longitude, or azimuthal) and $W'$ is the latitudinal gradient of vorticity.\\
This dispersion relation shows the important property that Rossby waves can propagate only in one direction, or equivalently produce vortices of only one orientation, cyclonic or anticyclonic, depending on the sign of $W'$. Differential rotation results in  a Doppler shift of the dispersion relation \citep{T01}, giving in a disk:
\begin{equation}
\omega-m\Omega(r)=\frac{mW'/r}{k_{r}^{2}+m^{2}/r^{2}}
\label{eq:disp}
\end{equation}
where $m$ is the azimuthal wavenumber. This applies in unmagnetized, isothermal disks of constant surface density \mt{ (so that $W$ appears here rather than \ls) whereas}, as shown in the above-mentioned works, the vorticity gradient $W'$ that appears in the numerator must include additional terms in more complex conditions (gradients of density, giving \ls, and magnetic field strength, giving \lb, polytropic equation of state, etc.).
\item The fact that $\Omega$ depends on $r$, while as discussed above the right-hand side of equation \ref{eq:disp} has a definite sign (choosing $m$ positive for definiteness) means that, depending on the sign of $W'$ (or the gradient of  \ls\ or \lb\ depending on the physics described), a Rossby wave can propagate only on one side of its corotation radius, defined as the radius in the disk where the left-hand side of equation \ref{eq:disp} vanishes. Since in general the energy of waves is positive beyond corotation, and negative inside it, this also means that the wave can have only either positive or negative energy, contrary to density waves that  can propagate on both sides of corotation and have accordingly positive or negative energy.
\item The dispersion relation also shows that Rossby waves in disks propagate in an annulus starting at the corotation radius, contrary to spiral density waves which have a forbidden band centered at corotation but can propagate on both sides beyond this band. It was however shown \citep{T01} that a Rossby wave is usually sheared away by differential rotation before it has completed even one oscillation cycle, unless some mechanism (such as the instability mechanisms discussed here) regenerates it continuously.
\end{itemize}
Most disk instability mechanisms rely on an exchange of energy (and the associated angular momentum) between waves of positive and negative energy: this is the case, for instance, of the Swing amplification of spiral waves in galaxies (where waves are driven by self-gravity), which also applies to the Papaloizou-Pringle instability \citep{PP85} (where they are driven only by the gas pressure), and also of the MHD instability first described by \citet{THSP90} where they are driven by the Lorentz force. In these three cases a spiral wave propagating inside corotation can be amplified by coupling to another spiral propagating beyond corotation, with which it exchanges energy and angular momentum. Amplification results from the fact that the wave inside corotation can increase its amplitude by transfering energy and momentum to the wave beyond corotation. The spirals can also couple to Rossby waves near corotation. This process, known as corotation resonance, is stabilizing or destabilizing depending on the gradient of \ls\, which decides whether the Rossby wave has a positive or negative energy. \mt{Given the form of \ls,  this is usually stabilizing for spiral modes in unmagnetized disks \citep{LBK72, Mark76, PP87, NGG87,PL89}, unless the radial density gradient is unusually strong. On the other hand, since the gradients act differently on \lb, it  has been found to give rise to the Accretion-Ejection Instability in disks threaded by a vertical magnetic field \citep{TP99}.}\\
Modes amplified by the Swing or by the corotation resonance depend on the existence of an inner disk edge where the spiral waves can be reflected. This allows them to travel back and forth between the inner edge and the corotation radius, forming standing patterns (\ie\ normal modes) when their frequency is such that they verify an integral phase condition.\\
The situation changes however when \ls\ (or \lb) has an extremum in the disk. Let us first consider the case where it has a minimum: then according to the dispersion relation (\ref{eq:disp}) a Rossby wave whose frequency lies precisely at this minimum can propagate both beyond it, where \ls' is positive, and within it, where it is negative. The above-mentioned works have shown that this allows normal modes, \ie\ standing wave patterns trapped near the minimum, and that these modes are unstable, \ie\ grow exponentially by exchanging energy between their positive and negative energy components. When \ls\ has a maximum, one also finds similar unstable modes, with now a corotation radius offset from the maximum  of \ls: these modes are localized where Rossby waves can propagate, \ie\ between the corotation radius and the maximum of \ls. This instability mechanism was used recently by \cite{TM06} to explain the quasi-periodic behavior observed during the flares of Sgr A*, by the capture in the disk of a fragment of stellar wind which creates at a few to a few tens of gravitational radii an extremum of density and magnetic field, and thus of \lb.\\
The RWI thus establishes a standing vortex in the corotation region. Since it essentially exchanges gas (and magnetic flux) from inside corotation with gas from outside, it might be viewed as a form of the classical MHD interchange instability. However here the main energy source driving the instability is differential rotation (\ie\ ultimately gravitation), much stronger than the magnetic field gradient  responsible for the interchange mode, unless the disk becomes magnetically (rather than centrifugally) supported against gravity.
\subsection{Diskoseismologic modes}\label{sec:disko}
\begin{figure}[t] 
\plotone{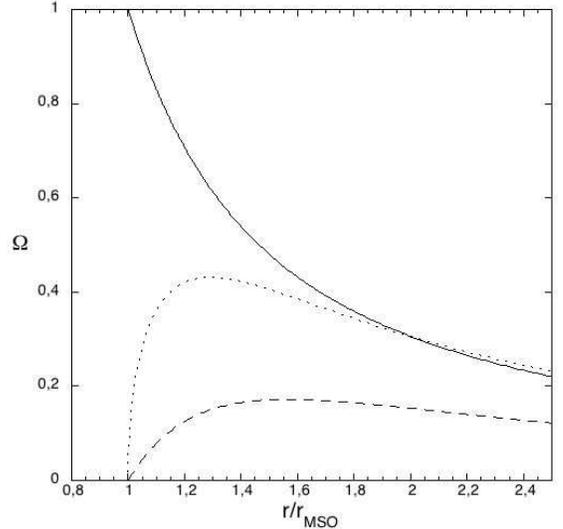}
\caption{Rotation frequency $\Omega$ (solid), epicyclic frequency $\kappa$ (dots) and $W=\kappa^{2}/2\Omega$ (dashed) in the Paczinsky-Witta pseudo-Newtonian potential.}
\label{fig:rotcurve}
\end{figure}

In all of these earlier applications, the extremum of $\cal L$ was mostly considered as due to 
an extremum in the radial density profile. However, starting from a very different point of 
view --- the seismology of relativistic accretion disks--- \cite{NW91} described modes associated 
with the same extremum of $\cal L$, but due this time to the behavior of $\kappa$: specifically,
the fact that the relativistic rotation curve in the disk of a compact object is characterized 
by the existence of a Marginally Stable Orbit (MSO), where $\kappa$ vanishes. Inside this orbit 
(at $r_{MSO}=6GM/c^2$ for a Schwarzschild black hole), $\kappa$ is imaginary, meaning that the 
motion of fluid particles is unstable and they rapidly spiral toward the compact object. Moving toward larger radii beyond the MSO 
$\kappa$ first increases, but then decreases again since
$\kappa \approx\Omega\sim r^{-3/2}$ in the Keplerian portion of the disk. \mt{Thus $\kappa$ and $W$ have maxima,} and usually so does $\cal L$, as shown in figure \ref{fig:rotcurve}, slightly beyond the MSO.\\
%
%
The non-axisymmetric g-modes of \citeauthor{NW91}, which are trapped in this extremum, are thus similar to the RWI, but it was not realized at that time that this similarity made them unstable (and very strongly, as will be seen below) by the same  mechanism found in the above-mentioned works. In what follows we will use the definition of \cite{LGN03}, who classify as g-modes the ones for which `most of the action takes place near corotation', \ie\ the RWI (whereas the AEI would be classified as a p-mode). In that work it was shown that the non-axisymmetric g modes which have nodes in their vertical structure are always stable. We are interested here in the fundamental g-mode, the one without nodes, \ie\ the one whose structure is essentially constant across the vertical extension of the disk. This allows us to consider an infinitely thin disk in vacuum \citep{TPC92}, as was the case \eg\ for galactic spiral modes \citep{GLB65, SK91} before 3D simulations were possible. \mt{On the other hand, this approach forbids us to consider the Magneto-Rotational Instability \citep[MRI,][]{BH91}, which depends on variations in the vertical direction: technically, when one studies the vertical structure of the modes in a disk with a realistic density stratification, one finds a discrete set of modes with $n_{z}=0,\ 1,\ 2,\ 3...$ nodes (this can be replaced by a vertical wavenumber $k_z$ when $n_{z}$ becomes large. The modes we consider correspond to $n_{z}=0$ \citep{TPC92}, whereas the MRI corresponds to $n_{z}\ge1$. Only fully three-dimensional simulations with the proper magnetic topology would permit to accomodate both types of instabilities.}
\\ 
\mt{\section{Numerical results}\label{sec:Num}}
In the present work we are mostly interested in the MHD form of the RWI, in a disk threaded by a vertical magnetic field, for a number of reasons: one is that accretion disks are magnetized, and that accretion from larger radii to the inner region should rely on the Magneto-Rotational Instability \citep{BH91}. A second reason is that in general we find that magnetic stresses make the modes significantly more unstable than in the unmagnetized case. \\
A last but key reason is that in previous work \citep{VT02} dedicated to the AEI we have derived a very important property of Rossby waves in that configuration: they represent a vortical motion applied to the fluid, and thus also to the footpoints of the magnetic field lines threading the disk. In an unmagnetized disk the linear physics would stop here, and one would have to turn to non-linearities to consider the ultimate evolution of the energy stored in the Rossby vortices, presumably when the perturbed motions become comparable with the sound speed \citep{NGG87}. On the other hand, if the disk is threaded by a vertical magnetic field, this vortical motion will propagate along the field lines into the corona of the disk as Alfv\'en (or more precisely Rossby-Alfv\'en) waves. Our result showed that these waves could carry to the corona a significant fraction of the accretion energy extracted from the disk. This provides naturally a channel to energize the corona and maybe form a jet or a wind. We will return to this point in our final discussion.\\
Since the physics and structure of the modes is essentially similar to the unmagnetized case we will not repeat here the linear theory. We will present in the next section numerical results, first solving the linearized set of equations for the unstable modes, and then turning to fully non-linear MHD disk simulations.
\subsection{Linear analysis}\label{sec:linear}
We first use the same procedure as in \cite{TP99}, to which we refer for a more detailed description: we write the linearized MHD equations for perturbations in the disk, and project them on a radial grid of $n_{r}$ points equally spaced in $s=\ln r$. \mt{The magnetic field in the vacuum surrounding the disk is described by a magnetic potential, whose source is the currents in the disk. The equilibrium magnetic field is taken normal to the disk. }
 \\
For simplicity, since the gas pressure plays little role, we use an isothermal equation of state, and at the MSO we use a reflecting boundary condition, $v_{r}=0$. We defer a discussion of this boundary condition to the next section, where this will be compared with other boundary conditions. At the outer radial boundary we use a radiation condition (\ie\ waves are allowed to freely escape), implemented as in \cite{TP99} by solving on a complex $s$ axis. This also improves the numerical convergence in the corotation region. \\
The only difference is thus that here we use the pseudo-newtonian potential of Paczinsky-Witta:
%
\begin{figure}[t] 
\plottwo{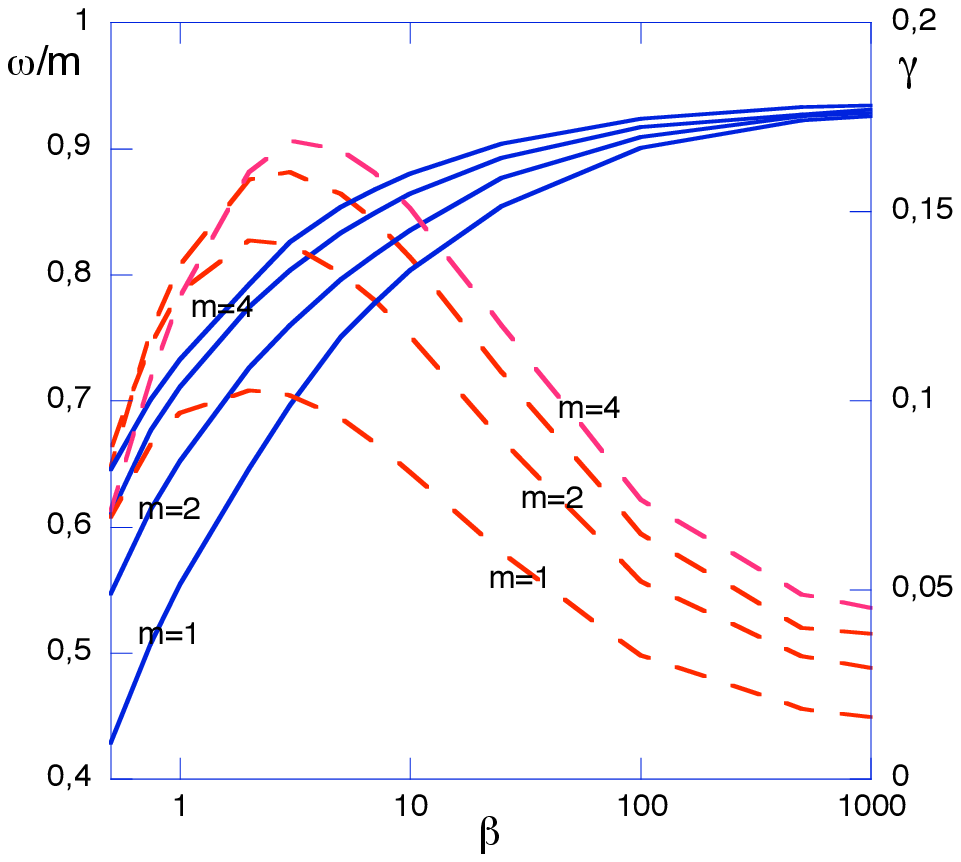}{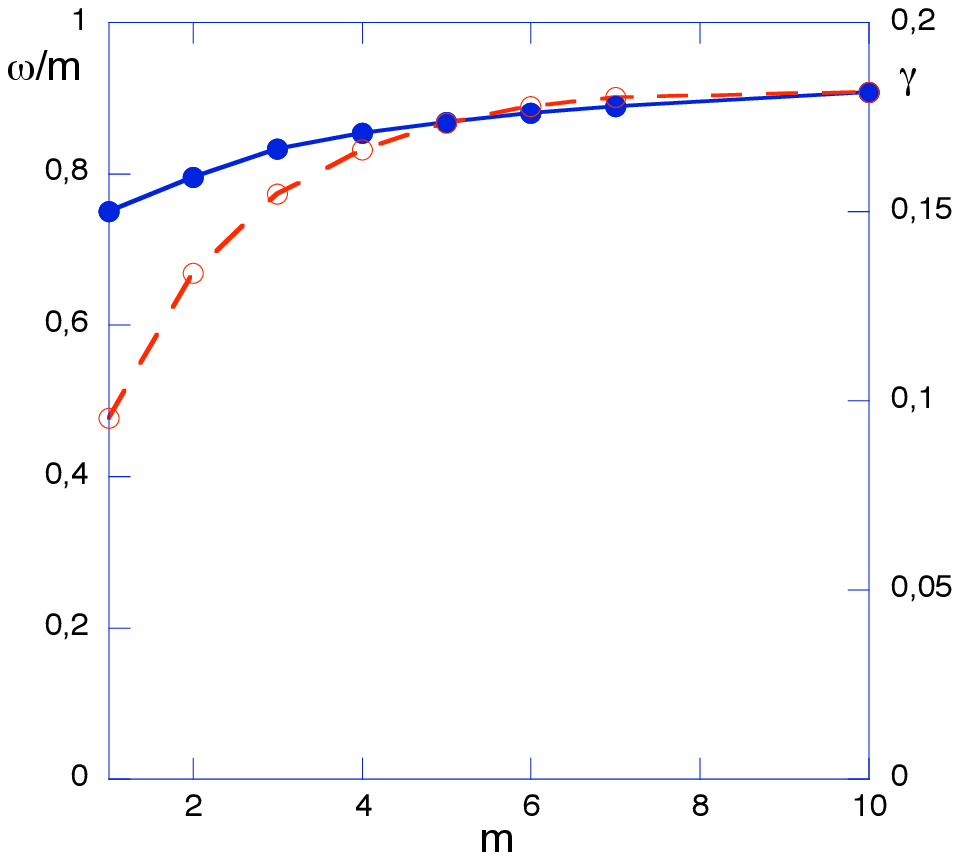}
\caption{Pattern speed $\omega/m$  (solid) and growth rate $\gamma$ (dashed) of the linear eigenmodes. Top, as a function of $\beta=8\pi p/B^{2}$; bottom, as a function of $m$ for $\beta=5$. The results are normalized to the rotation frequency at the MSO.}
\label{fig:linear}
\end{figure}
%
\begin{equation}
\Phi(r)=\frac{GM}{r-r_{G}}\;.
\end{equation}
This mimicks the relativistic rotation curve by allowing for the existence of an MSO, where $\kappa$ vanishes, at $r_{MSO}=3r_{S}\equiv 6GM/c^2$ for a Schwarzschild black hole. Figure \ref{fig:rotcurve} shows the radial profiles of $\Omega$ and $\kappa$ in this potential. %
The resulting discretized equations form a $(4n_{r}\times4n_{r})$ matrix, whose eigenvectors and eigenvalues give respectively the spatial structure of the perturbed quantities and the corresponding frequencies. Physically, these eigenmodes are standing wave patterns that have an exponential variation with time, defined by a real frequency (giving an azimuthal pattern speed $\omega/m$) and a growth rate. Among the $4n_{r}$ eigenvalues one easily separates by various means (essentially a convergence study) those of numerical origin, that result from the discretization, from the physical spectrum, in which we consider only the fundamental, most unstable mode. \\
For the results presented here we use simple power-law profiles for the surface density and magnetic field, $\Sigma \sim r^{-1/2}$ and $B \sim r^{-1}$, and a constant temperature. The only free parameters are then the sound velocity $c_{s}$ and the magnetization parameter $\beta=8\pi p/B^{2}$. We take varying values of $\beta$, and $c_{s}=.05\ r\Omega(r_{MSO})$, permitting a clear separation of scales with the rotation velocity. The behaviors we obtain are very weakly sensitive to the value of $c_{S}$, since thermal pressure plays only a minor role in the physics of the modes we consider (on the other hand decreasing $c_{S}$ would decrease the wavelength of the most unstable mode, in the unmagnetized case). Since the physics is very localized near the extremum of \lb, different profiles make very little practical difference.\\
Figure \ref{fig:linear} shows the frequency and growth rates of modes as a function of $m$ for $\beta=5$, and as a function of $\beta$ for $m=1$ to $4$. These results show that
\begin{itemize}
\item as expected, the modes have very close azimuthal pattern speeds ($\omega/m$), \mt{that correspond} to corotation radii very near the inner radius. They will thus appear very near multiples of a fundamental frequency. One has to be cautious about its precise value, since it depends on our assumption (modeling the relativistic rotation curve by a pseudo-newtonian potential) and also on the spin of the black hole.
\item although the instability already exists when $\beta$ becomes very large (in the limit $\beta\rightarrow \infty$ one gets  the hydrodynamical case of Lovelace and coworkers), one gets much higher growth rates ast $\beta$ approaches 1 (equipartition between the gas pressure and magnetic field pressure).
\item the growth rate of the $m=1$ mode is always 
smaller than for $m>1$. Modes with $m>1$, on the other hand, have very similar growth rates, so that we can expect minute differences in the disk physics, not included here, or in non-linear effects to select any value of $m\ge 2$ as the mode that will dominate oscillations of the disk. This readily makes this an excellent candidate to generate the observed HF-QPO at frequencies 2, 3, and sometimes higher multiples of an unobserved fundamental frequency.
\item The growth rates we find are very strong, giving exponentiation time of the order of the rotation period. This is due to the very strong gradient of \lb\ near the inner edge.
\end{itemize}
In the next section, we present the results of MHD simulations that will help us explore further these questions.
\subsection{Numerical simulations}\label{sec:sim}
We perform numerical simulations with the code presented in \cite{CT01}, to which we refer for a detailed description. As for the linear computations of section \ref{sec:linear}, it studies the evolution of an infinitely thin disk in vacuum, and the magnetic field outside the disk is derived from a magnetic potential related to the vertical component of the field at the disk by a Poisson equation. This makes the problem very similar to simulations of flat, self-gravitating disks and allows us to use methods (the Poisson kernel of
\citet{BT87}, and the FARGO algorithm of  \cite{M00}) developed in that context. The code
uses  a  2D cylindrical
($r,\phi$) grid, logarithmically spaced in radius to allow a good resolution in the inner disk region, while the outer boundary is kept far enough to avoid the influence of waves reflected there. 
A conservative scheme of the type described by \citet{SN92} is used to advance the fluid properties in
time. The resolution is typically taken to be 256x128 cells in $r$ and $\phi$
although higher resolution tests have been made for convergence studies. The main changes to the code of \cite{CT01} are the use of the pseudo-newtonian potential discussed above, and of various boundary conditions at the inner edge of the disk,  as presented in the following sections. \\
\subsection{Free flow boundary condition}\label{sec:plunging}
\begin{figure}[t] 
\plotone{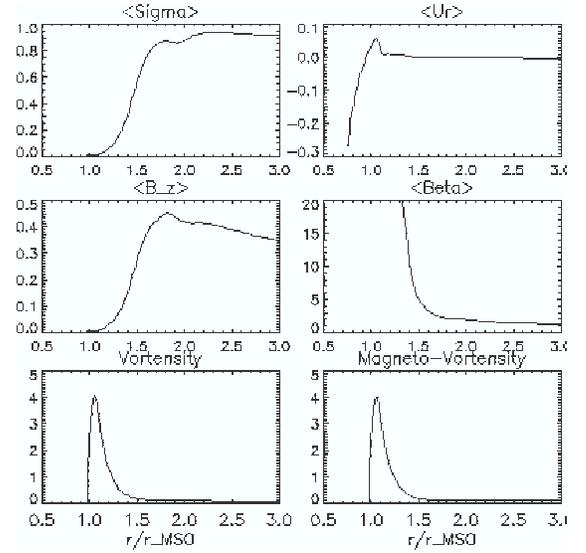}
\caption{Azimuthally averaged profiles of surface density, radial velocity, magnetic field, $\beta$, \ls\ and \lb\ in our first simulation, with a free flow boundary condition at $\approx .6 r_{MSO}$. Radii are scaled to 
$r_{MSO}$. Only the inner region is shown, although the simulation extends to $15 r_{MSO}$.
}
\label{fig:transpall}
\end{figure}

We first present simulations that allow the gas, once it has passed the MSO, to flow freely toward the black hole: it thus forms a `plunging region', as seen in 3D simulations that use the same hypothesis \citep{HK01, MM03}. This is implemented by fixing the inner grid point at typically $r=.6\ r_{MSO}$, with 30 grid points between this and the MSO, and allowing the gas to flow inward at this boundary. A plunging region is established self-consistently in $\sim$ one rotation time, and doesn't vary much afterward. Typically a sonic point is passed smoothly inward from the MSO, and the radial velocity at the inner point is $\sim 5c_{S}$. This also creates an outgoing pulse of density propagating outward at the sound speed, so that a few more rotation times are needed to establish quasi-stationary profiles in the inner few MSO radii. The resulting radial profiles are shown in figure \ref{fig:transpall}.\\
%
We might use these profiles as input in the numerical procedure of section \ref{sec:linear} for a linear solution, but find it unnecessary since at very low amplitudes the code provides simultaneously and reliably a linear solution for modes at different $m$ values. Also, the maximum of \ls\ and \lb\ near the MSO keeps evolving as the inner disk edge is slowly eroded by accretion caused by the instability. We could use viscosity (representing small-scale turbulence, presumably due to the MRI) to provide a continuous inflow that would stop  this erosion, but this would require additional modeling which is unnecessary at this stage. In practice, we stop the simulation when non-linearities become important, since they would probably be strongly affected by our thin-disk assumption and by the limited physics in the code, and at the present stage we restrict our discussion to the linearized physics.\\
\begin{figure}[t] 
\plotone{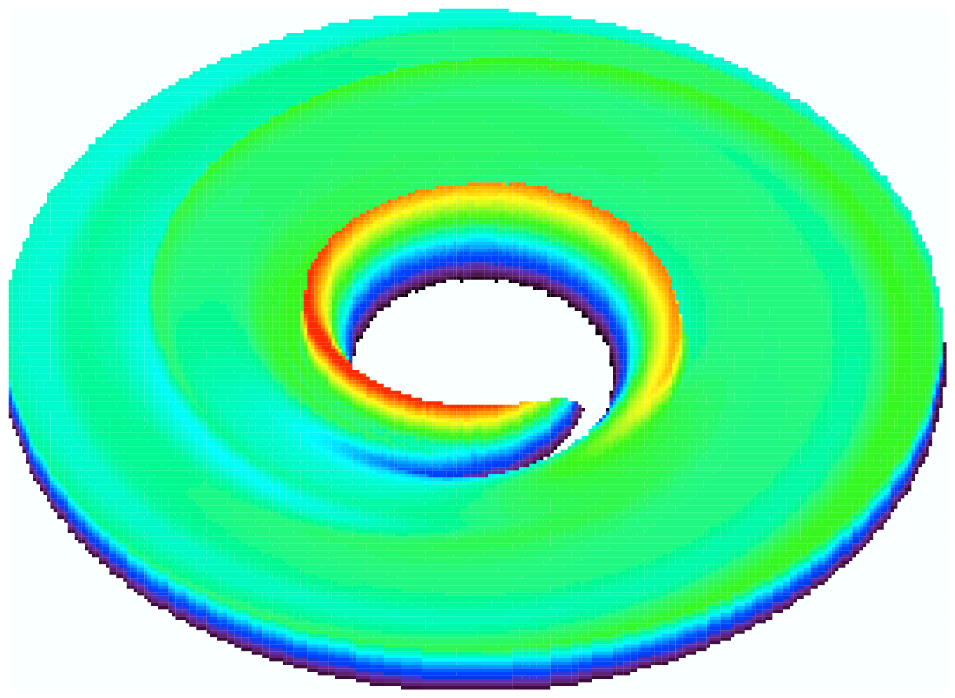}
\caption{Radial velocity  in our first simulation with free-flow boundary condition, showing the strong $m=1$ mode appearing as a one-armed spiral. for enhanced visibility the velocity is plotted both as a fake third dimension and as colors, and the disk is seen at an oblique angle.}
\label{fig:transpur}
\vskip .5cm
\plotone{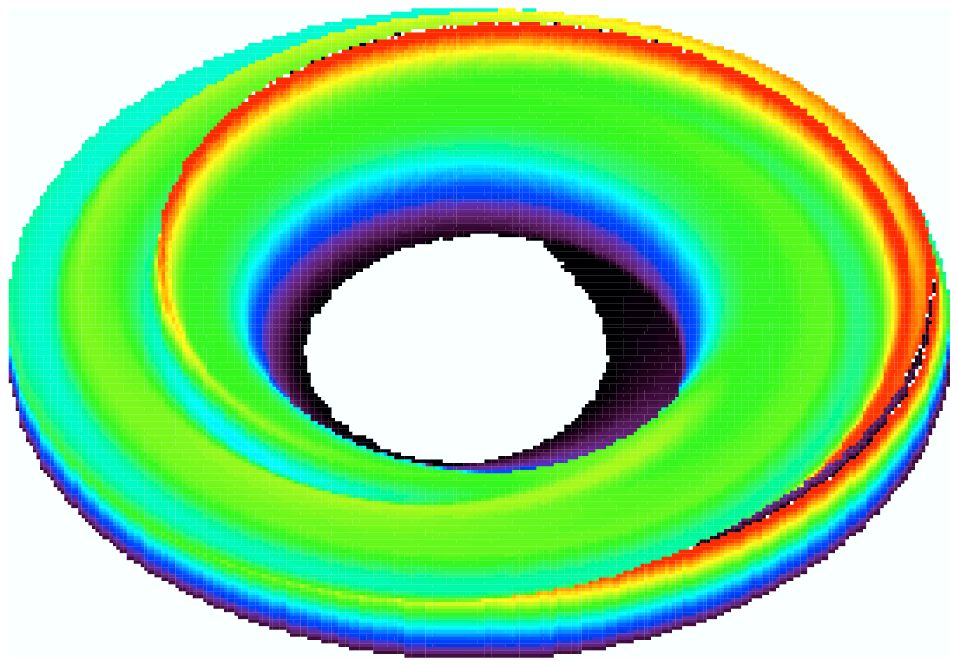}
\caption{Surface density in the disk.}
\label{fig:transprho}
\end{figure}
%
In these conditions, we do not expect modes, such as the AEI (or the Papaloizou-Pringle instability in the unmagnetized case) that depend on a reflective boundary \citep{BLA87}. On the other hand, as shown in figure \ref{fig:transpall}, the radial profile of $\kappa$ creates a sharp maximum of \lb\ just beyond the MSO, so that we can expect (and do obtain) a relativistic g-mode, unstable by the RWI mechanism, For all `reasonable' radial profiles of density and magnetic flux, and values of the physical parameters, we find that the $m=1$ mode is always dominant. Figures \ref{fig:transpur} and \ref{fig:transprho} show surface plots of the resulting radial velocity and surface density, showing the strong 1-armed spiral. Detailed analysis shows that this mode has a corotation  radius at $\approx\ 1.3\ r_{MSO}$, and is trapped in the  extremum of \lb\footnote{\mt{This is exactly where we expect it  given the profile of \lb: in the WKB limit, a Rossby wave can propagate inside corotation when the gradient of \lb\ is negative. Thus here it can propagate (in both radial directions, permitting the formation of a standing mode pattern) between the maximum of \lb\ at r=1.1 and the corotation radius at 1.3. If  \lb\ had a minimum, the corotation would be right at the minimum, and Rossby waves could propagate on both sides of it because of the inverse gradients of  \lb. This is for instance the case in \citet{TM06}, where a minimum of \lb\ is created by a maximum of density.}}. It is thus the expected non-axisymmetric mode of diskoseismology, unstable by the RWI mechanism. But simulations exploring a wide range of parameters and radial equilibrium profiles always show a strongly dominant $m=1$ mode, ruling out this mechanism as an explanation for the observed HF-QPO. 
\subsection{Reflecting boundary condition}\label{sec:bound}
On the other hand, this boundary condition that lets the gas and its magnetic flux flow freely to the black hole may not be the relevant one. This question actually reduces to whether or not the disk carries a non-vanishing vertical (poloidal) magnetic flux, and what happens to that flux when it has been advected to the inner disk edge. It has been argued \citep{LPP94} that turbulent resistivity might prevent the poloidal flux to be dragged with the gas in the disk. This relies however on rather crude assumptions and the result cannot be taken as granted. For instance, axisymmetric simulations of magnetized accretion-ejection structures \citep{CK02}, performed with hypothesis which should be favorable to this effect (turbulent magnetic diffusivity much stronger than viscosity), do show the poloidal flux advected with the gas. This is obtained in axisymmetric simulations, and thus does not even rely on the non-axisymmetric mechanism recently discussed by \citet{SU05}.\\
Most current three-dimensional MHD simulations of disk turbulence, in newtonian or general-relativistic gravity, consider disks with no net poloidal flux. This avoids a number of numerical and physical problems associated in particular with the boundary conditions, and with the initialization of the code. But these simulations have thus far found no QPO \mt{and, although they find jets from some form of the \citet{BZ77} mechanism, these jets are not collimated.\\
Most MHD models of \mt{collimated} jets powered by the disk (rather than the black hole spin) do require a net poloidal flux both to accelerate the gas and to collimate it. This is the case for centrifugally or magnetically driven jets \citep{BP82, LWS87, PP92}. It is also required for the AEI, which provides a possible explanation for the Low-Frequency QPO in that configuration. Since these models require a magnetic field of the order of equipartition ($\beta\sim 1$), it is likely that the poloidal flux that is established, in present 3D simulations, near the inner edge of the disk is too weak ($\beta\sim 10$) and localized to provide collimation}.
\\
If poloidal magnetic flux is advected with the gas, one has to consider that it must pile up in the central cavity between the disk and the event horizon: this leads us to discuss the black hole magnetospheric structure that holds this flux, involved in the General-Relativistic MHD processes of \cite{BZ77} and \cite{Pen69}. Although three-dimensional numerical models are available \citep[see \eg][]{DHK05}, they are still obtained in a configuration where the disk has no net magnetic flux. The most advanced axisymmetric simulation \citep{K05}, involving a Kerr Black Hole embedded in an external magnetic field, does show the Blandford-Znajek process at work, though apparently not efficiently enough to produce a relativistic jet.\\
This central magnetospheric structure can only be due to currents outside the Black Hole; it is commonly considered that they form a force-free structure in the magnetosphere, and a current ring which must sit at the inner edge of the disk since it must be tied to some mass in order to avoid rapid expansion.  This can be  seen for instance in \cite{K05} where a quasi-steady state is obtained after a few rotation times, and where the subsequent magnetic flux evolution is discussed in terms of  slow resistive processes. If this is the case \mt{one can expect} that, on the fast time scale of the modes we consider, this magnetospheric structure will oppose the penetration of the gas from the disk and thus effectively act as a reflecting boundary: it would oppose the fast flow associated with a plunging region, discussed in section \ref{sec:plunging}, or with the waves we consider here; but it would not preclude a gas flow on the slow time scales of reconnection and/or accretion. \\
%
%
\begin{figure}[t] 
\plotone{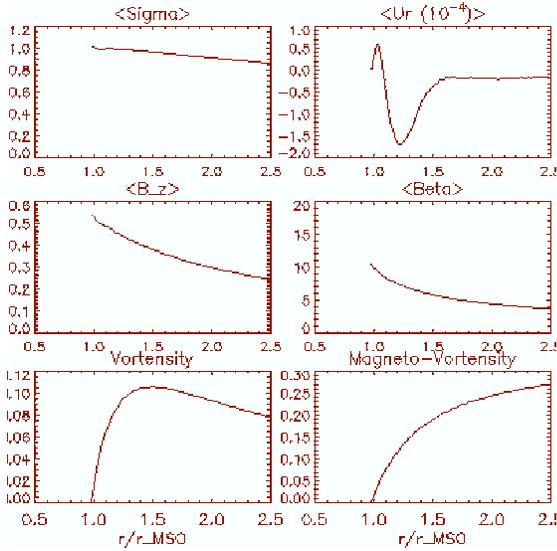}
\caption{Azimuthally averaged profiles of surface density, radial velocity, magnetic field, $\beta$, \ls\ and \lb\ in our second simulation, with a reflecting boundary condition at the MSO. Radii are scaled to $r_{MSO}$.}
\label{fig:reflectall}
\end{figure}
A complete solution of this problem is beyond the possibilities of present simulations, and we simply replace here this central magnetospheric structure by the same boundary condition (vanishing radial velocity at the inner radius) as in the linearized computation of section \ref{sec:linear}. \\
The most natural hypothesis on the location of this boundary (the disk/magnetosphere transition) is the MSO: both because this is where the gas dynamics anyhow undergoes a transition (more precisely, the inner disk radius cannot be smaller than the MSO without a very strong magnetic support, of the order of gravity rather than thermal pressure as assumed here), and because of observational results:  in the high states of microquasars, spectral fits indicate that the color radius consistently reaches a minimum and steady value. The relation between this and the actual inner radius of the disk must be taken carefully, since it is model dependent; but the evidence of this minimum value certainly tells us that the inner radius also reaches a minimal value, which in all likelihood must be the MSO.\\
The results of these simulations globally confirm the results of section \ref{sec:linear}:  starting from very low amplitude noise (from which the unstable modes grow), we can accurately measure the growth rates of the different $m$ modes, and find them in good agreement with the linear results, both for the radial structure of the perturbations and for the frequencies and growth rates.  Changing the initial noise component, or adding some viscosity to mimic the action of the MRI (supposed to act at smaller scale if we stay above its threshold $\beta=1$) allows the dominant mode to vary \mt{but the difference in growth rate with the $m=1$ mode is strong enough that only artificial initial conditions would allow $m=1$ to dominate up to the point where strong non-linearities lead us to stop the simulations.}
\\
\begin{figure}[t]
\plotone{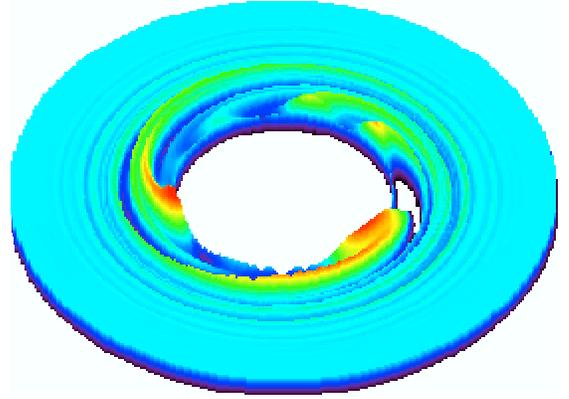}
\caption{Radial velocity  in our second simulation, showing a dominant $m=3$ mode.}
\label{fig:reflectur}
\end{figure}
\begin{figure}
\plotone{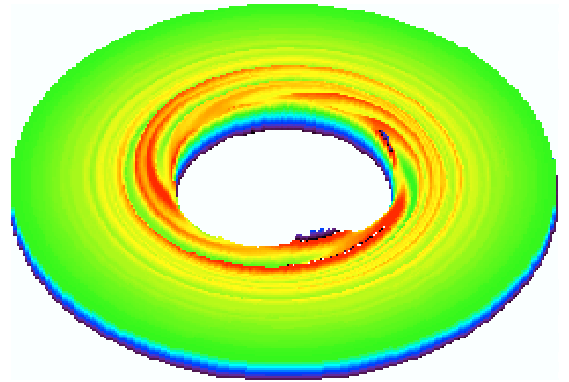}
\caption{Surface density in the disk.}
\label{fig:reflectrho}
\end{figure}
In agreement with the results of figure \ref{fig:linear}, the modes we obtain have their corotation at $\approx 1.1-1.2\ r_{MSO}$. Comparing with the radial profiles shown in figure \ref{fig:reflectall}, we see that this does not correspond to an extremum of \lb, but rather to a region where its slope is positive and large: this means that, rather than the RWI obtained with the free-flow boundary in section \ref{sec:plunging}, we are dealing here with an instability which is more of the AEI type \mt{(which relies on a reflecting boundary rather than on modes localized in the extremum of \lb)}, although the effect of the relativistic rotation curve forbids a clear separation between the two mechanisms. The strong slope of \lb, due to the relativistic rotation curve, explains the large growth rates we obtain.
\section{Discussion}\label{sec:disc}
In this paper we have put together the physics of diskoseismologic models, of the AEI, and of the RWI, to find that normal modes are strongly unstable in the accretion disk of black holes, especially when the disk magnetization is of the order of equipartition. The instability mechanism, relying on a coupling between spiral and Rossby waves in the disk, is the same that makes unstable the AEI and RWI, with variations depending on the conditions at the boundary between the disk and the black hole magnetosphere. \\
We find that, if the gas is allowed to flow freely across the Marginally Stable Orbit, the dominant mode is a one-armed ($m=1$) feature, not suitable to explain the high-frequency Quasi-Periodic Oscillations of microquasars. On the other hand we have argued that, \mt{if the disk carries a net poloidal flux which is advected toward the central region,}
the complex magnetospheric structure involved in the Penrose and Blandford-Znajek processes might act as a reflecting boundary for waves propagating in the disk.  \mt{Using this assumption}  we find, from linear analysis and numerical simulations, that all the modes with $m\ge 2$ have very similar pattern speeds ($\omega/m$) and growth rates, while the $m=1$ is less unstable. Small differences depending on additional physics (\eg\ three-dimensional effects, a realistic equation of state, viscosity from smaller-scale MRI turbulenceÉ) can select any of them, which would thus appear very near multiples of an unseen fundamental frequency, as the HF-QPO do. This would readily explain the 2:3 (and sometimes higher) frequency ratio observed between these QPO.\\
The modes are already unstable in non-magnetized disks, in which case they become forms of the RWI or Papaloizou-Pringle instabilities affected by the relativistic rotation curve. However they are much more unstable when the disk is threaded by a magnetic field approaching equipartition with the gas pressure ($\beta \ge 1$). The HF-QPO are observed in the Very High, or Steep-Power Law, state of the sources \citep[see \eg][]{McR06}, near the jet line in the color-color diagrams of \citet{FBK04}, 
except maybe in GRS 1915+105. This state is characterized by both a strong thermal disk emission and a steep power-law tail, probably from Comptonization in the corona. This contrasts with  the Low-Frequency QPO, seen in the low-hard state where the disk is hardly visible and the coronal emission is dominant, and that we consider as a manifestation of the AEI at $\beta \le 1$. These modes share the property that the accretion energy they extract from the disk is stored in a Rossby vortex, and we have shown \citep{VT02} that a substantial fraction of this energy can be emitted  toward the corona as an \alf wave. Energizing the corona by dissipation of this wave would thus  provide a natural explanation for the spectral characteristics of the sources when these QPO are observed.\\
Given these properties, it is natural to assume that the HF-QPO are seen at times where the magnetic field in the disk is below equipartition, allowing instability both at small scales by the MRI (explaining the thermal disk emission) and at large scale, in the vicinity of the MSO, by our instability mechanism. Further work should consider the relevance of this interpretation to the `Magnetic Floods' scenario \citep{TVRP04} we have recently presented for some of the cycles of GRS 1915+105. In this respect it is already remarkable that in this scenario, just as in the present work but from entirely different arguments, we were led to highlight the importance of the magnetic flux trapped between the inner disk edge and the event horizon of the black hole. \\
On the other hand, this work should lead to take with caution estimates of the black hole spin, derived from a tentative identification of the QPO frequency with the orbital frequency at the MSO, for a Kerr black hole whose mass is known. Here we find that the QPO frequencies are 2, 3, \ldots times a fundamental that is typically $\approx$~.8 times the MSO rotation frequency. This factor may change in fully relativistic computations, or in ones where additional physics such as three-dimensional effects, a realistic equation of state, viscous transport from small-scale turbulence, or the interaction with the central magnetospheric structure, would be taken into account, affecting in particular the radial profiles of density, temperature and magnetic field in the vicinity of the MSO. Much more elaborate simulations will be necessary to reliably assess these effects.
\acknowledgments
The authors express their gratitude to J. Rodriguez and R. Remillard for continuous and helpful discussions on the observed physics of the HFQPO.\\
PV is supported by NSF grants AST-9702484, AST-0098442, NASA
grant NAG5-8428, HST grant, DOE grant DE-FG02-00ER54600, the
Laboratory for Laser Energetics.
\bibliographystyle{apj}
\bibliography{ms} 

\begin{thebibliography}{21}
\expandafter\ifx\csname natexlab\endcsname\relax\def\natexlab#1{#1}\fi
%
\bibitem[Balbus \& Hawley(1991)]{BH91} Balbus, S.~A., \& 
Hawley, J.~F.\ 1991, \apj, 376, 214 

\bibitem[Binney \& Tremaine(1987)]{BT87} Binney, J., \& 
Tremaine, S.\ 1987, Princeton, NJ, Princeton University Press, 1987

\bibitem[Blaes(1987)]{BLA87} Blaes, O.~M.\ 1987, \mnras, 227, 
975 

\bibitem[Blandford \& Payne(1982)]{BP82} Blandford, R.~D., 
\& Payne, D.~G.\ 1982, \mnras, 199, 883 

\bibitem[Blandford \& Znajek(1977)]{BZ77} Blandford, R.~D., 
\& Znajek, R.~L.\ 1977, \mnras, 179, 433 

\bibitem[Casse \& Keppens(2002)]{CK02} Casse, F., \& 
Keppens, R.\ 2002, \apj, 581, 988 

\bibitem[{{Caunt} \& {Tagger}(2001)}]{CT01}
{Caunt}, S.~E., \& {Tagger}, M. 2001, \aap, 367, 1095

\bibitem[De Villiers et al.(2005)]{DHK05} De Villiers, J.-P., 
Hawley, J.~F., Krolik, J.~H., \& Hirose, S.\ 2005, \apj, 620, 878 

\bibitem[Drazin \& Reid(2004)]{DR81} Drazin, P.~G., \& Reid, W.~H.\ 2004, Hydrodynamic Stability.~ISBN 0521525411.~Cambridge, UK: Cambridge University Press, September 2004.

\bibitem[{{Dubrulle} \& {Zahn}(1991)}]{DZ91}
{Dubrulle}, B., \& {Zahn}, J.-P. 1991, Journal of Fluid Mechanics, 231, 561

\bibitem[Fender et al.(2004)]{FBK04} Fender, R.~P., Belloni, 
T.~M., \& Gallo, E.\ 2004, \mnras, 355, 1105 

\bibitem[Goldreich \& Lynden-Bell(1965)]{GLB65} Goldreich, 
P., \& Lynden-Bell, D.\ 1965, \mnras, 130, 125 

\bibitem[Hawley \& Krolik(2001)]{HK01} Hawley, J.~F., \& 
Krolik, J.~H.\ 2001, \apj, 548, 348 

\bibitem[Komissarov(2005)]{K05} Komissarov, S.~S.\ 2005, 
\mnras, 359, 801 

\bibitem[{{Li} {et~al.}(2001){Li}, {Colgate}, {Wendroff}, \& {Liska}}]{LCW01}
{Li}, H., {Colgate}, S.~A., {Wendroff}, B., \& {Liska}, R. 2001, \apj, 551, 874

\bibitem[{{Li} {et~al.}(2000){Li}, {Finn}, {Lovelace}, \& {Colgate}}]{LFL00}
{Li}, H., {Finn}, J.~M., {Lovelace}, R.~V.~E., \& {Colgate}, S.~A. 2000, \apj,
  533, 1023

\bibitem[{{Li} {et~al.}(2003){Li}, {Goodman}, \& {Narayan}}]{LGN03}
{Li}, L.-X., {Goodman}, J., \& {Narayan}, R. 2003, \apj, 593, 980

\bibitem[{{Lovelace} \& {Hohlfeld}(1978)}]{LH78}
{Lovelace}, R.~V.~E., \& {Hohlfeld}, R.~G. 1978, \apj, 221, 51

\bibitem[{Lovelace} {et~al.}(1999)]{LLC99}{Lovelace}, R.~V.~E., {Li}, H., {Colgate}, S.~A., \& {Nelson}, A.~F. 1999,
  \apj, 513, 805
  
\bibitem[Lovelace et al.(1987)]{LWS87} Lovelace, R.~V.~E., 
Wang, J.~C.~L., \& Sulkanen, M.~E.\ 1987, \apj, 315, 504 

\bibitem[Lubow et al.(1994)]{LPP94} Lubow, S.~H., Papaloizou, 
J.~C.~B., \& Pringle, J.~E.\ 1994, \mnras, 267, 235 

\bibitem[Lynden-Bell \& Kalnajs(1972)]{LBK72}Lynden-Bell, 
D., \& Kalnajs, A.~J.\ 1972, \mnras, 157, 1 

\bibitem[{{Machida} \& {Matsumoto}(2003)}]{MM03} {Machida}, M., \& {Matsumoto}, R. 2003, \apj, 585, 429

\bibitem[Mark(1976)]{Mark76} Mark, J.~W.~K.\ 1976, \apj, 205, 
363 

\bibitem[Masset(2000)]{M00} Masset, F.\ 2000, \aaps, 141, 
165 

\bibitem[McClintock \& Remillard(2006)]{McR06} McClintock, 
J.~E., \& Remillard, R.~A.\ 2006, Chapter 4 in "Compact Stellar X-ray Sources," eds. W.H.G. Lewin and M. van der Klis, Cambridge University Press, 2006

\bibitem[{{Narayan} {et~al.}(1987){Narayan}, {Goldreich}, \& {Goodman}}]{NGG87}
{Narayan}, R., {Goldreich}, P., \& {Goodman}, J. 1987, \mnras, 228, 1

\bibitem[{{Nowak} \& {Wagoner}(1991)}]{NW91}
{Nowak}, M.~A., \& {Wagoner}, R.~V. 1991, \apj, 378, 656

\bibitem[Papaloizou \& Lin(1989)]{PL89}Papaloizou, 
J.~C.~B., \& Lin, D.~N.~C.\ 1989, \apj, 344, 645 

\bibitem[{{Papaloizou} \& {Pringle}(1985)}]{PP85}
{Papaloizou}, J.~C.~B., \& {Pringle}, J.~E. 1985, \mnras, 213, 799

\bibitem[Papaloizou \& Pringle(1987)]{PP87}Papaloizou, 
J.~C.~B., \& Pringle, J.~E.\ 1987, \mnras, 225, 267 

\bibitem[Pelletier \& Pudritz(1992)]{PP92} Pelletier, G., \& 
Pudritz, R.~E.\ 1992, \apj, 394, 117 

\bibitem[Penrose(1969)]{Pen69}Penrose, R. 1969, Rev. Nuovo. Cim., 1, 252

\bibitem[Rodriguez et al.(2002)]{RVT02}Rodriguez, J., 
Varni{\`e}re, P., Tagger, M., \& Durouchoux, P.\ 2002, \aap, 387, 487 

\bibitem[{{Sellwood} \& {Kahn}(1991)}]{SK91}
{Sellwood}, J.~A., \& {Kahn}, F.~D. 1991, \mnras, 250, 278

\bibitem[Spruit \& Uzdensky(2005)]{SU05} Spruit, H.~C., \& 
Uzdensky, D.~A.\ 2005, \apj, 629, 960 

\bibitem[Stone \& Norman(1992)]{SN92} Stone, J.~M., \& 
Norman, M.~L.\ 1992, \apjs, 80, 753 

\bibitem[{{Tagger}(2001)}]{T01}{Tagger}, M. 2001, \aap, 380, 750

\bibitem[Tagger et al.(1990)]{THSP90}Tagger, M., Henriksen, 
R.~N., Sygnet, J.~F., \& Pellat, R.\ 1990, \apj, 353, 654 

\bibitem[Tagger \& Melia(2006)]{TM06} Tagger, M., \& Melia, 
F.\ 2006, \apjl, 636, L33 

\bibitem[{{Tagger} \& {Pellat}(1999)}]{TP99}
{Tagger}, M., \& {Pellat}, R. 1999, \aap, 349, 1003

\bibitem[Tagger et al.(1992)]{TPC92} Tagger, M., Pellat, R., 
\& Coroniti, F.~V.\ 1992, \apj, 393, 708 

\bibitem[Tagger et al.(2004)]{TVRP04}Tagger, M., 
Varni{\`e}re, P., Rodriguez, J., \& Pellat, R.\ 2004, \apj, 607, 410 

\bibitem[Varni{\`e}re et al.(2002)]{VRT02}Varni{\`e}re, P., 
Rodriguez, J., \& Tagger, M.\ 2002, \aap, 387, 497 
 
\bibitem[Varni{\`e}re \& Tagger(2002)]{VT02}Varni{\`e}re, 
P., \& Tagger, M.\ 2002, \aap, 394, 329 
 
\bibitem[Varni{\`e}re \& Tagger(2006)]{VT06}Varni{\`e}re, 
P., \& Tagger, M.\ 2006, \aap, 446, L13 

\end{thebibliography}





%

%













































\end{document}